\documentclass[a4paper,10pt]{article}
\usepackage[utf8]{inputenc}
\usepackage[numbers,compress,merge]{natbib}
\usepackage{hyperref}
\usepackage{graphicx}
\usepackage{amsmath,amssymb}
\usepackage{subfigure}
\usepackage{color,xcolor}
\usepackage[varg]{txfonts}
\usepackage[affil-it]{authblk}

\newcommand{\Ai}{A_{\omega l}^\text{in}}
\newcommand{\At}{A_{\omega l}^\text{tr}}
\newcommand{\Ar}{A_{\omega l}^\text{ref}}

\begin{document}
 
\title{Scalar scattering from charged black holes on the brane}

\author{Ednilton S. de Oliveira\footnote{ednilton@pq.cnpq.br}}
\affil{Faculdade de F\'isica, Universidade Federal do Par\'a,
	66075-110, Bel\'em, Par\'a, Brazil}

\date{\today}

\maketitle

\begin{abstract}
The differential scattering cross section of the massless scalar field localized on the 3-brane of charged static black holes in the ADD model is analyzed. While results valid in the entire range of the scattering angle can be obtained only via a numerical approach, analytical results can be obtained via the geodesic, Born, and glory approximations. The comparisons between numerical and analytical results lead to excellent agreements. The increase of the charge intensity has the consequence of increasing the width of the interference fringes in the scattering cross section. Its influence on the intensity of the scattered flux, however, depends on the dimensionality of the spacetime. Analyses for the special cases of uncharged and extremely charged black holes are included.
\end{abstract}


\section{Introduction}

The study of astrophysical objects has given an enormous step forward with the recent detections of gravitational waves~\citep{abbott2016prl116_061102,*abbott2016prl116_241103,*abbott2017prl118_221101}. These realizations mark the beginning  of a new epoch in science, when mankind started to look into the Universe in the perspective of gravitational waves. Also, these remarkable accomplishments  brought to our knowledge the existence of binary systems of stellar-mass black holes. This can be regarded as one more evidence that such objects populate the Universe in varied systems, with masses which go from some to million, or even billion, solar masses~\citep{begelman2003science300_1898}. On the other hand, electromagnetic radiation still gives the best possibilities of probing structures around black holes with sizes comparable to their event horizons~\citep{johannsen2015cqg33_113001,johannsen2015prl116_031101}. In particular, one of the black hole candidates which has attracted most attention of astrophysicists is Sgr A*, the black hole in the center of our galaxy, with estimated mass of $\sim 4 \times 10^6 M_\odot$.

The possibilities mentioned above have instigated some researches about how radiation emitted near, e.g. from the accretion disk~\citep{noble2007cqg24_s259}, and from far sources~\citep{bohn2015cqg32_065002} are scattered in the vicinity of black holes. When enough resolution is achieved, astrophysicists will finally observe the event horizon of these black holes as a shadow~\citep{falcke1999aj528_l13}, which can describe not only the characteristics of the black hole, as its rotation, but also input experimental constrains to alternative theories of gravity.

Although astrophysical black holes are undoubtedly the `labs' for testing the strong limit of gravity, it has also increased in the recent years the effort to observe black hole properties in labs. Such efforts can be put in two categories: one based on acoustic analogue systems~\citep{unruh1981prl46_1351,barcelo2005lrr8_12} and the other based on brane-world scenarios~\citep{arkani1998plb429_263,*antoniadis1998plb436_257,randall1999prl83_3370,cavaglia2002ijmpa18_1843,kanti2004ijmpa19_4899,emparan2008lrr11_6}. These two approaches are now in different levels; while acoustic systems have already been used to observe some of event horizon consequences, as the stimulated Hawking radiation~\citep{hawking1975cmp43_199,weinfurtner2010prl106_021302} and superradiance~\citep{torres2016arxiv1612.06180}, the consequences of the interaction between higher-dimensional black holes and 3-brane fields still remain in the theoretical level. The appearance of black holes in particle colliders, as the LHC (see Ref.~\citep{Park2012ppnp67_617} for a recent review), is one real possibility of studying black hole phenomenology in local experiments conditioned to one of the most fundamental problems in physics: the existence of extra dimensions\footnote{See Refs.~\citep{Sirunyan2017plb774_179,Aaboud2017arxiv1709.10440} for recent experimental constrains on black hole production at the LHC.}.

Semi-classical black holes created at the LHC would rapidly evaporate via Hawking radiation. This fact has instigated the study of scattering properties of higher-dimensional black holes, once Hawking radiation is directly related to the greybody factors which measure the probability of the black hole absorbing scattered waves. Works have been done considering mainly the Arkani-Hamed--Dimopoulos--Dvali (ADD) scenario, in which uncharged~\citep{emparan2000prl85_499,kanti2002prd66_024023,kanti2003prd67_104019,harris2003jhep10_014,jung2004jhep09_005,kanti2005prd71_104002,marinho2016arxiv1612_05604}, charged~\citep{jung2005npb717_272}, and rotating~\citep{Ida2003prd67_064025,Ida2005prd71_124039,Ida2006prd73_124022,Harris2005plb633_106,Duffy2005jhep09_049,Casals2005jhep02_051,Casals2006jhep03_019,creek2007prd75_084043} black holes have been analyzed. Some work about wave properties, as quasinormal modes, around black holes in the Randall-Sundrum scenario has been recently developed~\cite{toshmatov2016prd93_124017,molina2016prd93_124068}.

Although it is more likely that semi-classical black holes produced at the LHC, after a balding phase, should be rotating with a latter ``Schwarzschild phase''~\citep{Park2012ppnp67_617}, studying non-rotating charged black holes could help to foresee some of the features of such objects. This happens thanks to the fact that charged black holes share some properties\footnote{An example of similar behavior between charged and rotating black holes is the fact that they shrink if they get more charge or angular momentum. As a consequence, the scalar absorption cross section of charged~\citep{crispino2009prd79_064022} or rotating~\citep{Macedo2013prd88064033} black holes decrease if their respective charge or angular momentum increase.} with rotating black holes with the advantage of being static. Therefore, the scalar scattering of static charged black holes in the context of the ADD scenario represents a simple model which can help to understand not only the consequences of extra dimensions but also the physics of mini-black holes which can appear at the LHC in the near future.

In the present work we analyze the scattering properties of a small (static) charged black hole for the masseless scalar field restricted to the 3-brane in the ADD model. We compare results for black holes with different charge intensities, including the uncharged~\citep{marinho2016arxiv1612_05604} and extreme cases. In Sec.~\ref{sec:wave_scatt} we present the geometry of the analyzed system as well as the scattering properties of the massless scalar field in such geometry. In Sec.~\ref{sec:approx} we present the analytical methods used to obtain approximated results which can be compared with the numerical ones. The numerical results obtained via the partial-wave method are presented in Sec.~\ref{sec:results} together with their comparisons with the analytical results. We present our conclusions in Sec.~\ref{sec:conc}. Here we adopt the speed of light $c = 1$.

\section{Wave scattering}\label{sec:wave_scatt}

The spacetime of higher-dimensional non-rotating charged black holes was found by Myers and Perry~\citep{myers1986ap172_304}. This solution generalizes the Reissner-Nordström solution~\citep{chandra1983} to a $d$-dimensional spacetime and it is given by
\begin{equation}
ds^2 = f(r)dt^2 - f(r)^{-1} dr^2 - r^2 d\Omega_{n+1}^2,
\label{le_nd}
\end{equation}
where $n = d-3$ and
\begin{equation}
 f(r) = 1 - \frac{C}{r^n} + \frac{D^2}{r^{2n}},
 \label{f}
\end{equation}
with the constants related to the black hole mass $M$ and charge $Q$ by \cite{myers1986ap172_304}
\begin{equation}
 C = \frac{16\pi G_d M}{(n+1)\Omega_{n+1}},
 \label{eq:C}
\end{equation}
and
\begin{equation}
 D = \pm \left(\frac{8\pi G_d}{n(n+1)} \right)^{1/2} Q.
 \label{eq:D}
\end{equation}
Above, $d\Omega_{n+1}^2$ is the line element of a unit $(n+1)$-sphere and
$$ \Omega_{n+1} = \frac{2\pi^{\frac{n+2}{2}}}{\Gamma\left(\frac{n+2}{2}\right)}
$$
is its area.

The spacetime given by Eq.~\eqref{le_nd} describes a black hole if $D^* \equiv 2|D|/C \le 1$ and a naked singularity if $2|D|/C > 1$. In the first case, $f(r)$ possesses horizons located at
\begin{equation}
r_{\pm} = \left(C/2 \pm \sqrt{C^2/4 - D^2} \right)^{1/n},
\label{rhs}
\end{equation}
with $r_+$ being the event horizon.
Schwarzschild-Tangherlini black holes are obtained by letting $D \to 0$, in which case $r_- \to 0$. In the limit $|D| \to C/2 $ we have an extreme black hole. In this case $r_- \to r_+$ and we have only one horizon. 

Here we study small charged black holes on the 3-brane described by the ADD model~\citep{arkani1998plb429_263,antoniadis1998plb436_257}. Such black holes interact with particles localized on the 3-brane via the following geometry~\citep{emparan2000prl85_499,jung2005npb717_272}\footnote{Following Ref.~\citep{emparan2000prl85_499} we assume that the black hole is formed from matter on the brane, which has negligible self-gravity and thickness.}:
\begin{equation}
 ds^2 = f(r)dt^2 - f(r)^{-1} dr^2 - r^2 d\Omega_2^2,
 \label{brane}
\end{equation}
with $f(r)$ given by Eq.~\eqref{f}, where $d\Omega_2^2$ represents the line element of a 2-sphere of unitary radius.

The dynamics of the massless scalar field is governed by the Klein-Gordon equation:
\begin{equation}
 \frac{1}{\sqrt{-g}} \partial_{\mu} \left ( \sqrt{-g} g^{\mu\nu} \partial_\nu \Phi \right)  = 0,
 \label{KG}
\end{equation}
where the metric $g^{\mu\nu}$ is implicitly defined in Eq.~\eqref{brane}, and $g$ is its determinant.

A complete understanding of the scattering properties of a system requires the dynamic equations to be fully solved. However, the scattering problems of only a few systems have complete analytic solutions and one usually has to apply approximated and/or numeric methods~\citep{gottfried_etal-2004}. The situation is similar in the context of black hole scattering~\cite{futterman_etal1988}. In such cases, the main line adopted to solve the problem consists in separating the wave into partial waves with different angular momenta. Using the so-called ``partial-wave method'' requires a separation of variables. In the present case, we use the spherical symmetry of the spacetime to expand $\Phi$ in terms of partial waves proportional to the scalar spherical harmonics $Y_l^m(\theta,\phi)$, i.e., $\Phi_{\omega lm} = [\psi_{\omega l}(r)/r] Y_l^m(\theta,\phi) e^{-i\omega t}$. By doing so, the radial function $\psi_{\omega l}$ can be shown to satisfy the following equation:
\begin{equation}
 f \frac{d}{dr}\left(f \frac{d\psi_{\omega l}}{dr} \right) + \left[\omega^2 - V_l(r) \right ] \psi_{\omega l} = 0,
 \label{radial_eq}
\end{equation}
where the effective potential is given by
\begin{equation}
 V_l(r) = f\left[\frac{f'}{r} + \frac{l(l+1)}{r^2} \right],
 \label{V}
\end{equation}
with the prime standing for differentiation with respect to $r$.

The effective potential tends to zero in two regions: (I) near the black hole horizon and (II) at infinity\footnote{Note that for $r \gg L$, whith $L$ being the size of the extra dimensions, the metric will recover its 4-dimensional form, $f \sim 1-2G_4 M/r+\mathcal{O}(1/r^2)$~\citep{emparan2000prl85_499}. However, for black holes much smaller than the size of the extra dimensions, the main correction $2G_4M/r$ is already very small at $r \sim L$ and the scattered wave will not be significantly modified in the region between $r \sim L$ and $r \to \infty$. Therefore, we consider that the phase shifts, and therefore the scattering amplitude, will not be significantly modified far from the black hole by the correction $2G_4M/r$.} Therefore, we can obtain approximated solutions for $\psi_{\omega l}$ in such regions. In order to do so, we introduce the tortoise coordinate defined as
\begin{equation}
 \frac{d}{dr_*} = f \frac{d}{dr},
 \label{tortoise}
\end{equation}
so that the radial equation can be written as
\begin{equation}
 \frac{d^2 \psi_{\omega l}}{dr^2_*} + \left[\omega^2 - V_l(r_*) \right ] \psi_{\omega l} = 0,
 \label{radial_eq-tor}
\end{equation}
from where we can directly see that, for the scattering problem
\begin{equation}
 \psi_{\omega l} \sim \left\{
 \begin{array}{lr}
  \At e^{-i\omega r_*} & (\text{region I});\\
  \Ai e^{-i\omega r_*} + \Ar e^{i\omega r_*} & (\text{region II}).
 \end{array} \right.
 \label{asymp}
\end{equation}
The coefficients $\Ai$, $\Ar$, $\At$ are related to the incident, reflected and absorbed quantities of each partial wave. Here we consider a planar monochromatic wave impinging upon the black hole. By considering the initial wave as $\Phi_\text{inc} \propto e^{-i\omega z}$ we automatically remove the $\phi$-dependence of the wave, once it is spinless. With such considerations, the scattering amplitude can be given in terms of partial waves by~\cite{futterman_etal1988}:
\begin{equation}
 f_\omega(\theta) = \frac{1}{2i\omega} \sum\limits_{l=0}^{\infty} (2l+1) \left[ e^{2i\delta_l(\omega)} - 1\right] P_l (\cos\theta),
 \label{amp}
\end{equation}
where $P_l(\cdot)$ are the Legendre polynomials and the phase shifts, $\delta_l (\omega)$, are defined by
\begin{equation}
 e^{2i\delta_l(\omega)} = (-1)^{l+1} \Ar/\Ai.
 \label{ps}
\end{equation}

The differential scattering cross section follows directly from the scattering amplitude:
\begin{equation}
 \frac{d\sigma_\text{el}}{d\Omega} = |f_\omega (\theta)|^2.
 \label{dscs}
\end{equation}

\section{Approximations}\label{sec:approx}

Some approximations are useful to test the precision of our results. These are the case of the geodesic limit and the glory approximation. Other approximations can be used together with the numerical computation to improve the precision of our results in certain limits. This is the case of the Born approximation. We present such approximation methods below and compare them with the partial-wave-method results in Sec.~\ref{sec:results}.

\subsection{High frequency}\label{sec:geo}

At high frequencies we can use the geodesic approach to describe the cross sections. Since we are working with massless particles, we have to consider null geodesics. From Eq.~\eqref{brane} we obtain:
\begin{equation}
 \dot{s}^2 = f(r) \dot{t}^2 - f(r)^{-1} \dot{r}^2 - r^2 (\dot{\theta}^2 + \sin^2\theta \dot{\phi}^2) = 0,
 \label{geo}
\end{equation}
where the dot denotes differentiation with respect to an affine parameter. Once again we can use the spherical symmetry of the problem to eliminate the $\theta$-dependence of the motion by making $\theta = \pi/2$. Also, because of the symmetries of the spacetime, we have two motion constants
\begin{equation}
 E = f \dot{t},
 \label{E}
\end{equation}
and
\begin{equation}
 L = r^2 \dot{\phi},
\end{equation}
which are related to the energy and the angular momentum of the particle, respectively.
Through these constants we can define the impact parameter as $b = L/E$. By doing so, after some manipulation, Eq.~\eqref{geo} results in
\begin{equation}
 \left(\frac{du}{d\phi} \right)^2 = h_b(u) \equiv \frac{1}{b^2} - u^2f(1/u),
 \label{eq_phi}
\end{equation}
where we have made the change of variable $r \to u = 1/r$. We can also take the derivative of~\eqref{eq_phi} to generate a second-order differential equation. This leads to
\begin{equation}
 \frac{d^2u}{d\phi^2} + u = \left(\frac{n}{2}+1\right)Cu^{n+1} 
 - (n+1)D^2u^{2n+1}.
 \label{eq_phi2}
\end{equation}
By doing $d^2u/d\phi^2 = 0$ we can obtain the radius of the unstable orbit of the black hole for massless particles, $r_c = 1/u_c$. Inserting this value in $h_b(u)$ at Eq.~\eqref{eq_phi} and equaling it to zero, we obtain the critical parameter $b_c = r_c/[f(r_c)]^{1/2}$. Particles which start their motion at infinity with $b < b_c$ will be absorbed by the black hole, while the ones with $b >  b_c$ will be scattered back to infinity.

If we choose $b > b_c$, the smallest real root of $h_b(u)$ will describe the returning point of the geodesic, $u_0 = 1/r_0$. We are interested in computing the deflection angle $\Theta(b)$ which from~\eqref{eq_phi} can be given by:
\begin{equation}
 \Theta(b) = 2 \int\limits_0^{u_0} h_b(u)^{-1/2} du-\pi.
 \label{def_ang}
\end{equation}
The integration in Eq.~\eqref{def_ang} has known solutions in some cases, as in the case of 4-dimensional Schwarzschild ($n=1$, $D=0$)~\citep{darwin1959prsla249_180}, Reissner-Nordström ($n=1$, $|D| > 0 $)~\citep{crispino2009prd79_064022}, and the canonical acoustic black hole ($n=4$, $D=0$)~\citep{dolan2009prd79_064014}~\footnote{The effective metric which describes the canonical acoustic hole~\citep{visser1997cqg15_1767} is the same as the one of 7-dimensional Schwarzschild black holes induced on the 3-brane. However, the physics of the two systems in intrinsically different; the curvature around a canonical acoustic hole is experienced only by acoustic perturbations in the fluid flow which forms the hole~\cite{unruh1981prl46_1351}.} generally written in terms of elliptic integrals. However, we cannot find a general expression for Eq.~\eqref{def_ang} here. Instead, we deal with it numerically in order to obtain results for the classical scattering cross section.

The scattering angle relates to the deflection angle as $\theta = |2m\pi-\Theta|$, where $m$ are the number of times the geodesic circles the black hole. The classical scattering cross section can be given by
\begin{equation}
 \frac{d\sigma_\text{el}^\text{(cl)}}{d\Omega} = \sum_b\frac{b}{\sin\theta} \left| \frac{db}{d\theta} \right|,
 \label{cl_scs}
\end{equation}
where the sum in $b$ takes into account the cases in which rays incoming with different $b$s are scattered in the same direction, i.e., same scattering angle but different deflections.

\begin{figure}[!htb]
\centering
\includegraphics[width=\textwidth]{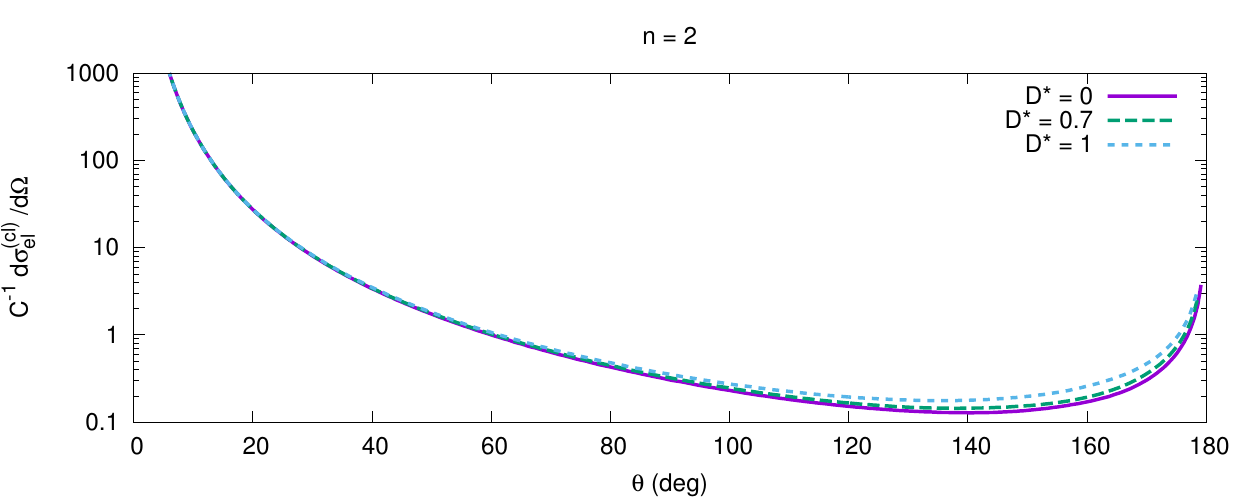}
\includegraphics[width=\textwidth]{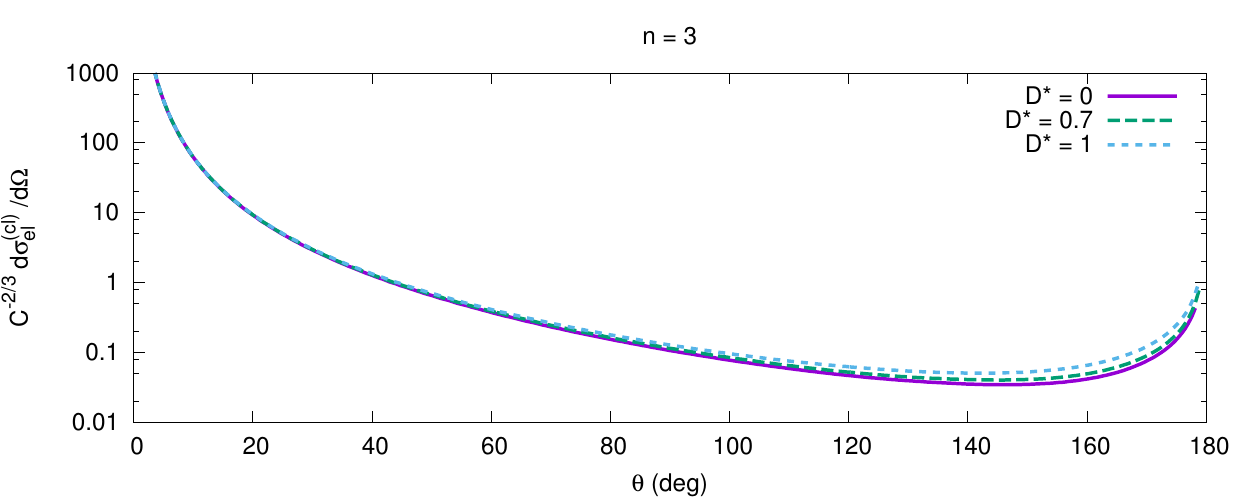}
\caption{Null-geodesic scattering cross section localized on the 3-brane of higher-dimensional charged black holes. Here we have results for the cases $n = 2$ (top) and $n = 3$ (bottom).}
\label{fig:geo_sca}
\end{figure}

In Fig.~\ref{fig:geo_sca} we present some results for the classical scattering cross section, Eq.~\eqref{cl_scs}. We have considered the cases $n = 2,3$ with $D^* = 0,0.7,1$. For fixed values of $n$ we see that the results are different in the near-backward direction but tend to be equal in the forward direction. This is an evidence that the charge plays a less important role as far from the black hole the particle passes. Also, the black holes scatters more as more charged they are since the differential scattering cross section is larger for higher values of $D^*$. Analogous results have been observed for other values of $n$ and also in the case of Bardeen black holes~\citep{macedo2015prd92_024012}.

\subsection{Glory approximation}\label{sec:glory}

Geodesic approximation usually applies well in the limit of small scattering angles. Another approximation which can be useful to compare with the partial-wave results is the \emph{glory} approximation~\citep{futterman_etal1988,matzner1985prd31_1869} valid in the limit $\theta \approx 180^\circ$. For spherically symmetric spacetimes, this approximation is given by the general formula:
\begin{equation}
 \frac{d\sigma_\text{el}^{(\text{gl})}}{d\Omega} = 2\pi\omega b_g^2 \left|\frac{db}{d\theta} \right|_{\theta = \pi} J_{2s}^2(b_g \omega \sin\theta),
 \label{glory}
\end{equation}
where $b_g$ is the impact parameter of rays which are scattered at $\theta = 180^\circ$, $s$ is the particle spin, and $J_\nu(\cdot)$ is the Bessel function of the first kind.

The values of $b_g^2 |db/d\theta |_{\theta = \pi}$ and $b_g$ will govern the intensity and the fringe widths which describe the interference pattern caused by rays scattered in opposite senses. These values vary according to the spacetime curvature, what makes each spacetime be identified by a different oscillatory pattern. The glory approximation~\eqref{glory} applies only if the scattered particle interferes with other particles without having its state altered. This is the case of most scattering processes around black holes, but it has a few exceptions, e.g., when helicity-reversing scattering takes place leading to non-zero backscattered electromagnetic radiation in Reissner-Nordström spacetimes~\citep{crispino2014prd90_064027}.

Since we cannot find a general form of the deflection angle for all $n$ and $D^*$, here we make numerical estimations for the cases which we compare with the results obtained via the partial-wave method in Sec.~\ref{sec:results}. Such estimations are listed in Tab.~\ref{tab:gl_params}. From this table we can infer that $b_g$ tends to decrease with the increase of $D^*$, while $|db/d\theta|$ tends to increase. Therefore, we can already predict that the fringes of interference will be wider as $D^*$ increases for fixed $n$s or when $n$ increases keeping the value of $D^*$ unaltered. We cannot say much about the intensity of the peaks, since it is proportional to $b_g^2 |db/d\theta| $. It has been shown in Ref.~\citep{crispino2009prd79_064022} that the glory intensity does not obey a monotonic behavior for case $n = 1$, decreasing as the black hole charge intensity increases at first, but increasing again as $D^* \to 1$.

\begin{table}
\centering
$n =  2$\\
 \begin{tabular}{cccc}
 \hline\hline
 $D^*$ & 0 & 0.7 & 1 \\\hline
 $C^{-1/2}b_g$ & 2.01 & 1.94 & 1.86 \\\hline
 $C^{-1/2}|db/d\theta|$ & $1.29\times 10^{-2}$ & $1.56\times 10^{-2}$ & $2.16\times 10^{-2}$ \\
 \end{tabular}\\
 \vspace{0.5cm}
 $n = 3$\\
 \begin{tabular}{cccc}
  \hline\hline
  $D^*$ & 0 & 0.7 & 1 \\\hline
  $C^{-1/3}b_g$ & 1.75 & 1.72 & 1.68 \\\hline
  $C^{-1/3}|db/d\theta|$ & $3.48\times 10^{-3}$ & $4.27\times 10^{-3}$ & $5.90\times 10^{-3}$ \\\hline
 \end{tabular}\\
 \vspace{0.5cm}
 $ n = 4$\\
 \begin{tabular}{cccc}
 \hline\hline
  $D^*$ & 0 & 0.7 & 1 \\\hline
  $C^{-1/4}b_g$ & 1.61 & 1.59 & 1.57 \\\hline
  $C^{-1/4}|db/d\theta|$ & $1.22\times10^{-3}$ & $1.50\times 10^{-3}$ & $2.03\times 10^{-3} $\\\hline
  
 \end{tabular}
 \caption{Glory parameters for $n = 2$ (top), $n=3$ (middle), and $n=4$ (bottom) considering the cases $D^* = 0, 0.7, 1$.}
 \label{tab:gl_params}
\end{table}

\subsection{Born approximation}\label{sec:born}

Via the Born approximation~\citep{gottfried_etal-2004}, it can be shown that in the weak-field limit~\cite{marinho2016arxiv1612_05604}:
\begin{equation}
 \delta_l^{(B)} \approx \frac{\sqrt{\pi}}{2(n-1)} \frac{\Gamma\left(\frac{n+3}{2}\right)}
 {\Gamma\left(\frac{n+2}{2}\right)} \frac{\omega^n C}{(l+1/2)^{n-1}}. \qquad( n > 1).
 \label{born}
\end{equation}
Although the formula above was obtained considering Schwarzschild black holes, it can be applied for charged black holes as well. This is justified by the fact that the Born approximation applied in the scattering from black holes is usually valid in the weak-field limit. However, in such limit, the most relevant term in the interaction between the black hole and neutral particles comes from the black hole mass term, leading its charge to be important only in a second order. This is already clear in the 4-dimensional case, where the first order of the deflection angle in the weak-field regime depends only on the black hole mass, with the charge appearing only in the second-order term~\citep{eiroa2002prd66_024010,bhadra2003prd67_103009,sereno2003prd69_023002}.

\begin{figure}[!htb]
\centering
\includegraphics[width=\columnwidth]{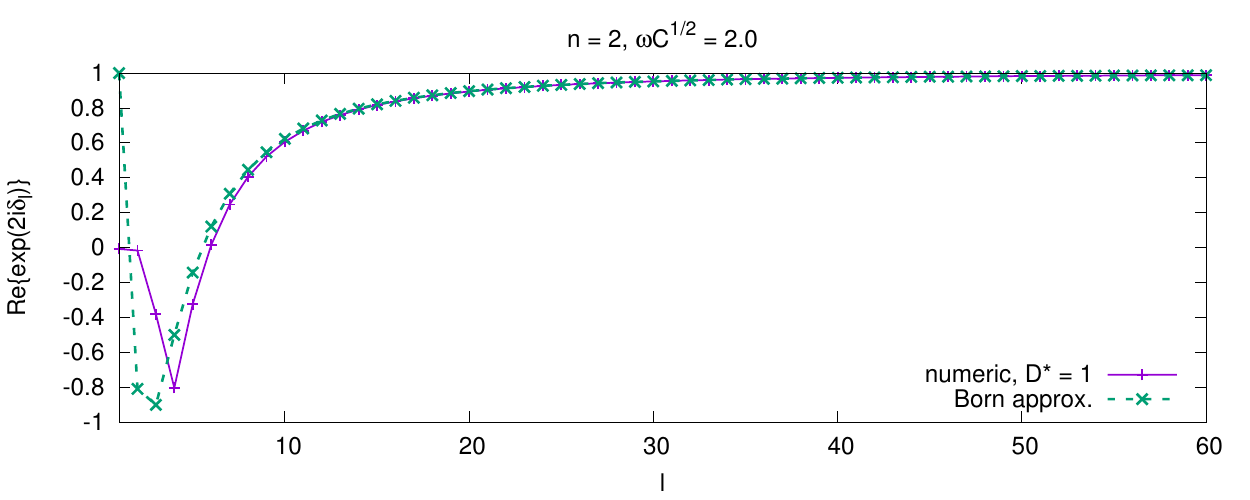}
\includegraphics[width=\columnwidth]{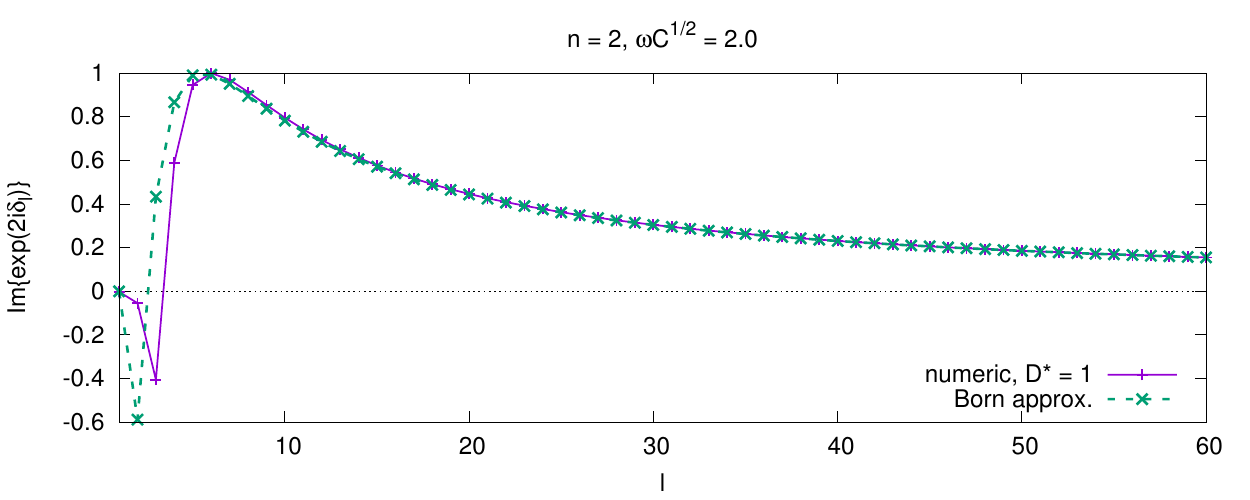}
\caption{Comparison of phase shifts obtained numerically ($D^* = 1$) and via Born approximation, Eq.~\eqref{born}, for $n = 2$ and $\omega C^{1/2} = 2.0$.}
\label{fig:comp_born_ps}
\end{figure}

In order to make such analysis more quantitative, in Fig.~\ref{fig:comp_born_ps} we compare the phase shifts obtained numerically in the case of extremely charged black holes for $n = 2$, $\omega C^{1/2} = 2.0$ and the Born approximation, Eq.~\eqref{born}. It is visible that the these results agree very well already for $l \sim 20$. For larger values of $n$ the spacetime becomes asymptotically flat even faster. Therefore, we expect that the agreement between the numerical phase shifts and the Born approximation to be very good for smaller values of $l$.

An important prediction can be done considering Eq.~\eqref{born}. For the cases which the approximation applies to, the scattering amplitude can be separated into two terms:
\begin{equation}
f_\omega (\theta) = f_\omega^{(\text{num})} (\theta) + f_\omega^\text{(ana)} (\theta),
\label{f_split}
\end{equation}
where
\begin{equation}
f_\omega^{(\text{num})} (\theta) \equiv \frac{1}{2i\omega} \sum\limits_{l=0}^{l_m} (2l+1) \left[ e^{2i\delta_l ^{(\text{num})}(\omega)} - 1\right] P_l (\cos\theta),
\label{f-num}
\end{equation}
with $\delta_l^{(\text{num})}$ obtained numerically, and
\begin{equation}
f_\omega^{(\text{ana})} (\theta) \equiv \frac{1}{2i\omega} \sum\limits_{l_m+1}^{\infty} (2l+1) \left[ e^{2i\delta_l ^{(B)}(\omega)} - 1\right] P_l (\cos\theta),
\label{f-ana}
\end{equation}
with $l_m$ within the regime of validity of the Born approximation, which is $l \gg \omega C^{1/n}$. Once $\delta_l^{(B)} \ll 1$, we can write:
\begin{eqnarray}
f_\omega^{(\text{ana})} (\theta) & \approx & \frac{1}{\omega} \sum\limits_{l_m+1}^{\infty} (2l+1) \delta_l ^{(B)}(\omega) P_l (\cos\theta) \nonumber \\
& \approx & \frac{\sqrt{\pi} \Gamma\left(\frac{n+3}{2}\right) \omega^{n-1}C}{(n-1)\Gamma\left(\frac{n+2}{2}\right)} \sum\limits_{l_m+1}^{\infty} (l+1/2)^{2-n} P_l (\cos\theta).
\label{f-ana2}
\end{eqnarray}

The largest contributions from the Legendre polynomials to the sum of Eq.~\eqref{f-ana2} happen in the forward direction, where $P_l(1) = 1$, so that the terms in the series  will depend on $(l+1/2)^{2-n}$ in the forward direction. Therefore, the scattering amplitude will be finite for $n \ge 4$. We use expression~\eqref{f_split} in order to obtain the differential scattering cross section in cases $n \ge 4$. However, we have to truncate the series at a point where the sum of the remaining terms can be neglected.

\section{Results}
\label{sec:results}

In this section we present the results for the cross section obtained numerically via the partial-wave method. The method consists basically in computing the phase shifts \eqref{ps} by comparing the numerical solution of the radial equation~\eqref{radial_eq} with the asymptotic solutions~\eqref{asymp}. The sum in the scattering amplitude, Eq.~\eqref{amp}, is developed through two different approaches, depending on the value of $n$. For $n \le 3$, we use phase shifts obtained numerically together with a method of reduced series described in Ref.~\citep{yennie1954pr85_500} since the scattering amplitude sum converges purely in such cases. Therefore, this method guarantees that we will obtain convergent scattering amplitudes considering a relative small number of phase shifts~\citep{dolan2006prd74_064005}. For $n > 3$, the cases in which the differential scattering cross section is finite in the entire range of $\theta$~\citep{marinho2016arxiv1612_05604}, we split the scattering amplitude into two terms. The first part is computed with the sum of terms which include phase shifts obtained numerically until $l_m$. The remaining part of the series is then computed with phase shifts obtained via Born approximation, Eq.~\eqref{born}, from $l_m + 1$ until $l_\text{max}$, with $l_m$ within the regime of validity of Eq.~\eqref{born}. This approach has been introduced to the study of black hole scattering in Ref.~\citep{dolan2009prd79_064014}.

\begin{figure*}
 \centering
 \includegraphics[width=0.49\columnwidth]{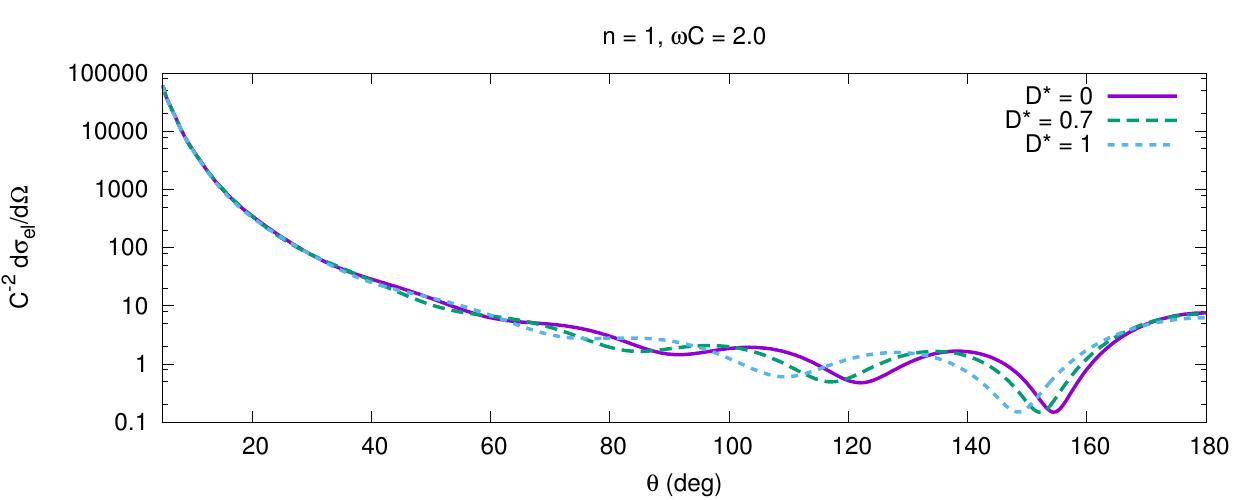}
 \includegraphics[width=0.49\columnwidth]{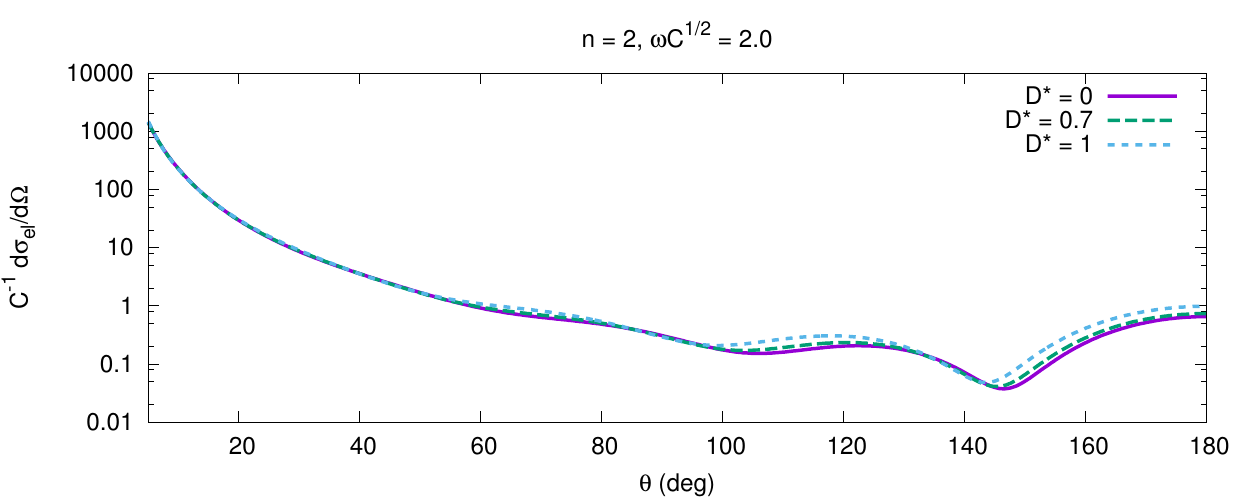}
 \includegraphics[width=0.49\columnwidth]{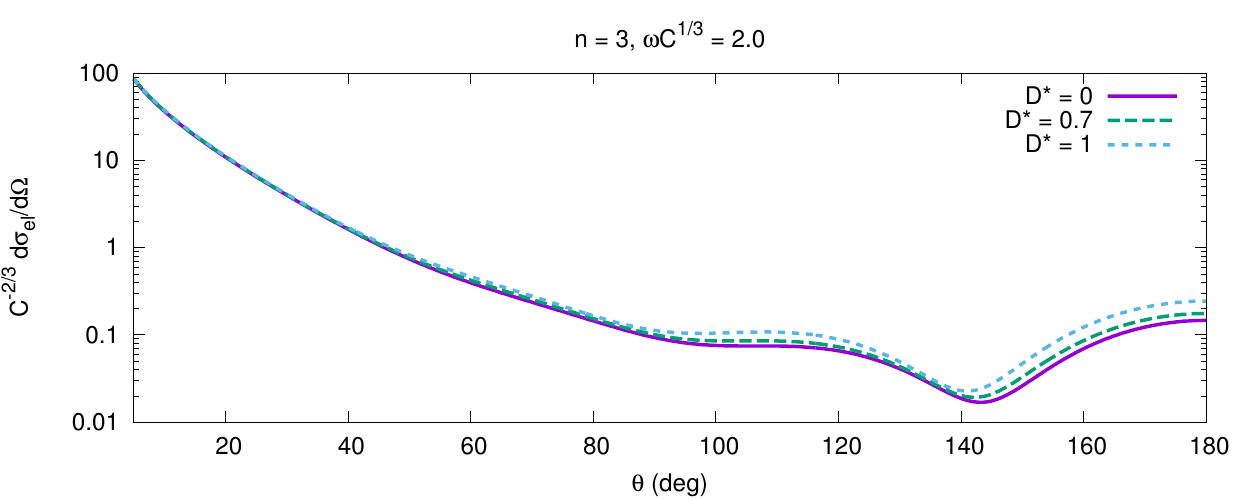}
 \includegraphics[width=0.49\columnwidth]{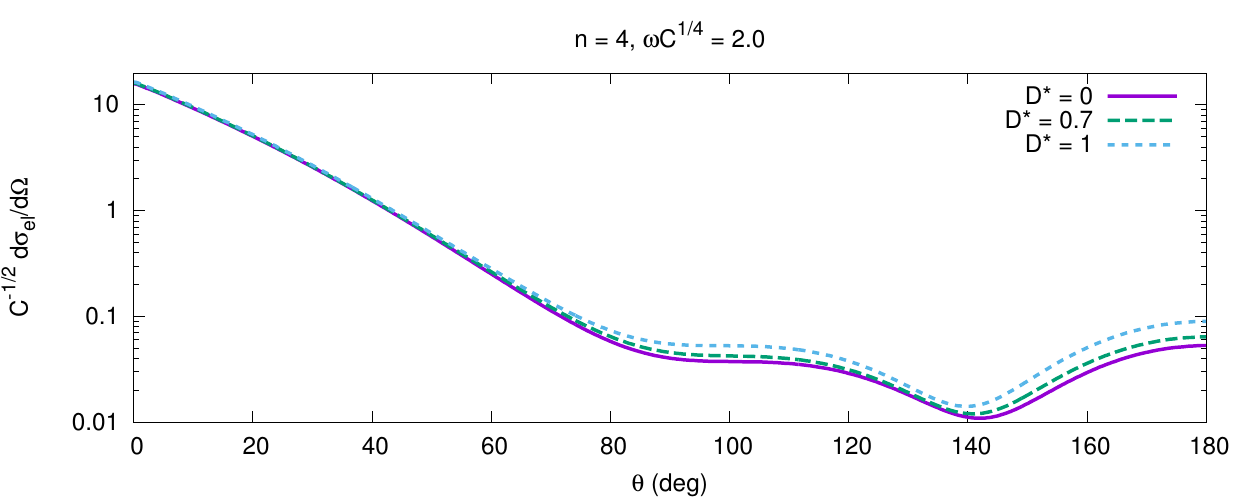}
 \caption{Scattering cross section of charged black holes for the massless scalar field on the 3-brane for $n = 1$ (top-left), $n = 2$ (top-right), $n = 3$ (bottom-left), and $n = 4$ (bottom-right). }
 \label{fig:charge_comp}
\end{figure*}

In Fig.~\ref{fig:charge_comp} we present the scalar differential scattering cross sections for charged black holes on the brane for $n = 1,2,3,4$, $D^* = 0,0.7,1$, and $\omega C^{1/n} = 2.0$. As we can see, the charge of the black hole plays an important role in the scattering for large values of $\theta$. In this regime, the widths of interference tend to widen with the increase of $D^*$, as anticipated in Tab.~\ref{tab:gl_params}, where we observed that $b_g$ decreases with the increase of $D^*$. We observe an increase in the intensity of the interference peaks in the cases $n > 1$, but not in the case $n = 1$, as already observed in Ref.~\citep{crispino2009prd79_064022}. For low values of $\theta$ the charge of the black hole has little influence in the scattering, and we can observe that the results for $D^* = 0,0.7,1$ for all values of $n$ presented tend to coincide in the regime $\theta \lesssim 40^\circ$.

\begin{figure}[!htb]
\centering
\includegraphics[width=0.49\textwidth]{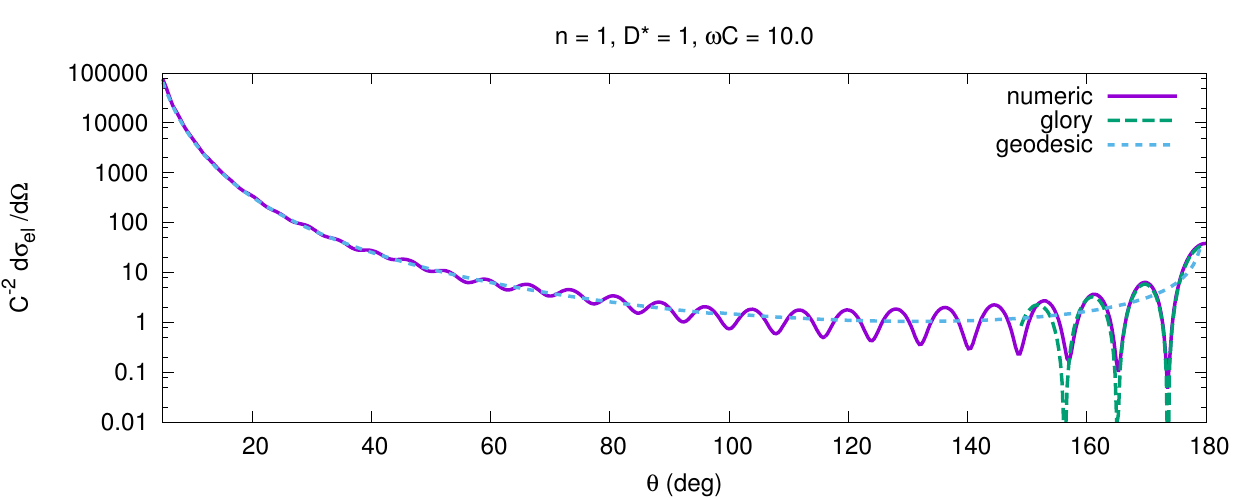}
\includegraphics[width=0.49\textwidth]{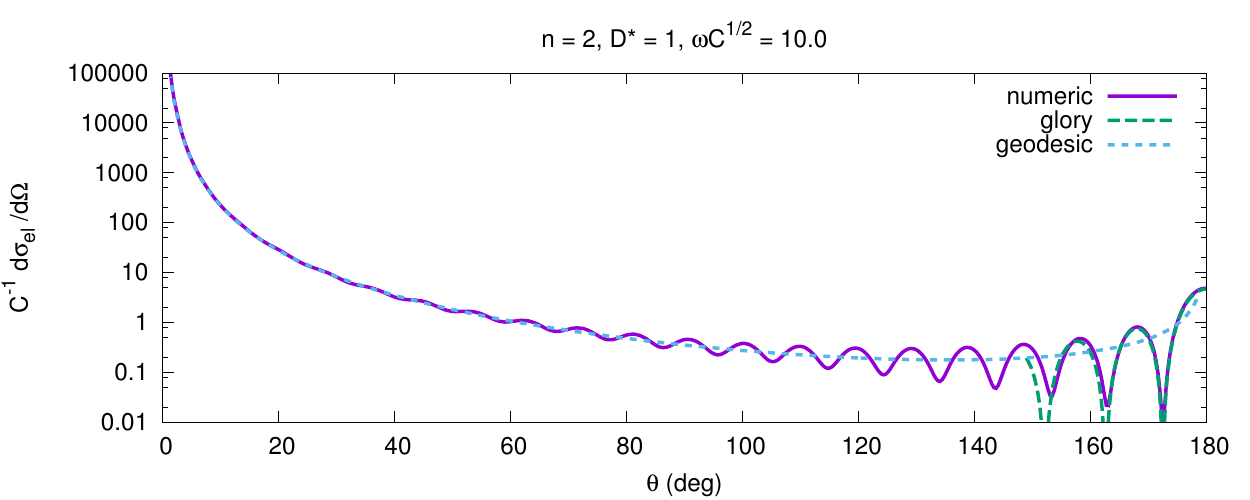}
\includegraphics[width=0.49\textwidth]{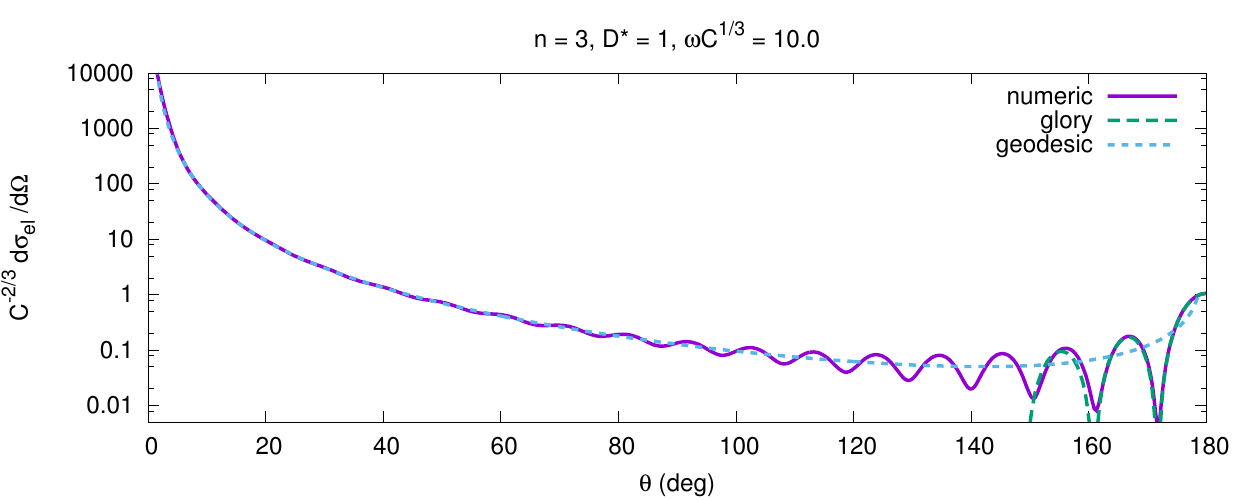}
\includegraphics[width=0.49\textwidth]{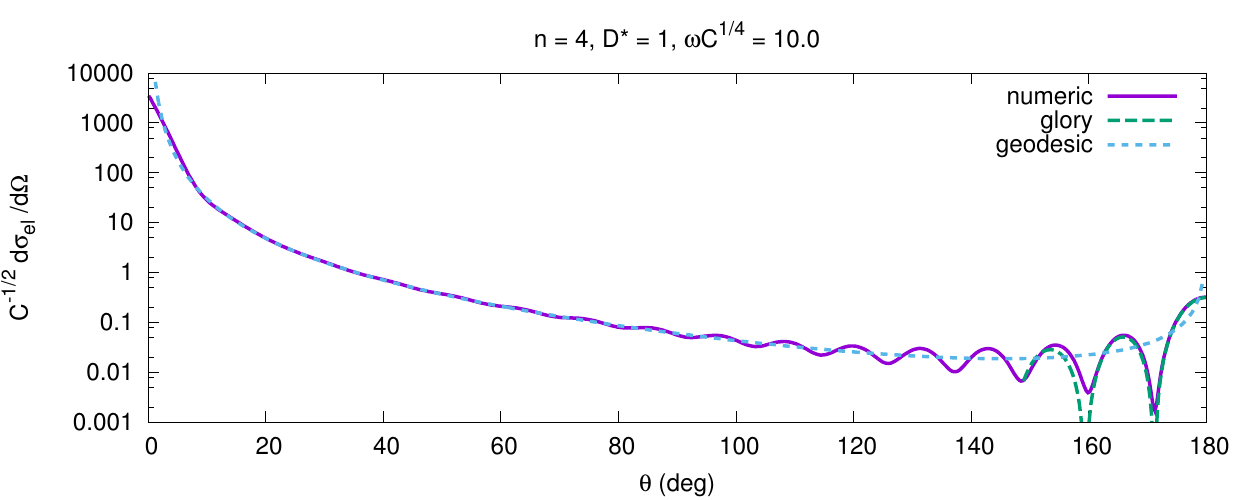}
\caption{Comparisons of the differential scattering cross section obtained via the partial-wave method and via the geodesic and glory approximations. Here we consider extreme black holes in all cases with $\omega C^{1/n} = 10.0$.}
\label{fig:comp}
\end{figure}

We compare the numerical results with the geodesic and the glory approximations, presented in Secs.~\ref{sec:geo} and~\ref{sec:glory} respectively, in Fig.~\ref{fig:comp}. There we consider a relative high value of frequency ($\omega C^{1/n} = 10.0$) once such approximations are valid in the high-frequency regime. We see that the agreement with the glory approximation is excellent in all cases for $\theta \gtrsim 160^\circ$. The geodesic approximation, however, is only a very good approximation in cases $n = 1, 2, 3$, but not in case $n=4$. In this case the geodesic method predicts that the flux scattered in the forward direction should be infinity while the wave analysis predicts a finite flux in the same direction. We expect the geodesic approximation being recovered only in the classical limit, i.e., $\omega C^{1/n} \to \infty$, for the cases in which $n \ge 4$. Although we presented only results for extreme black holes in Fig.~\ref{fig:comp}, we have observed similar agreements for the cases $D^* < 1$. The same comparison has been done for the case of uncharged black holes in Ref.~\citep{marinho2016arxiv1612_05604} with similar conclusions.

\begin{figure}[!htb]
\centering
\includegraphics[width=\columnwidth]{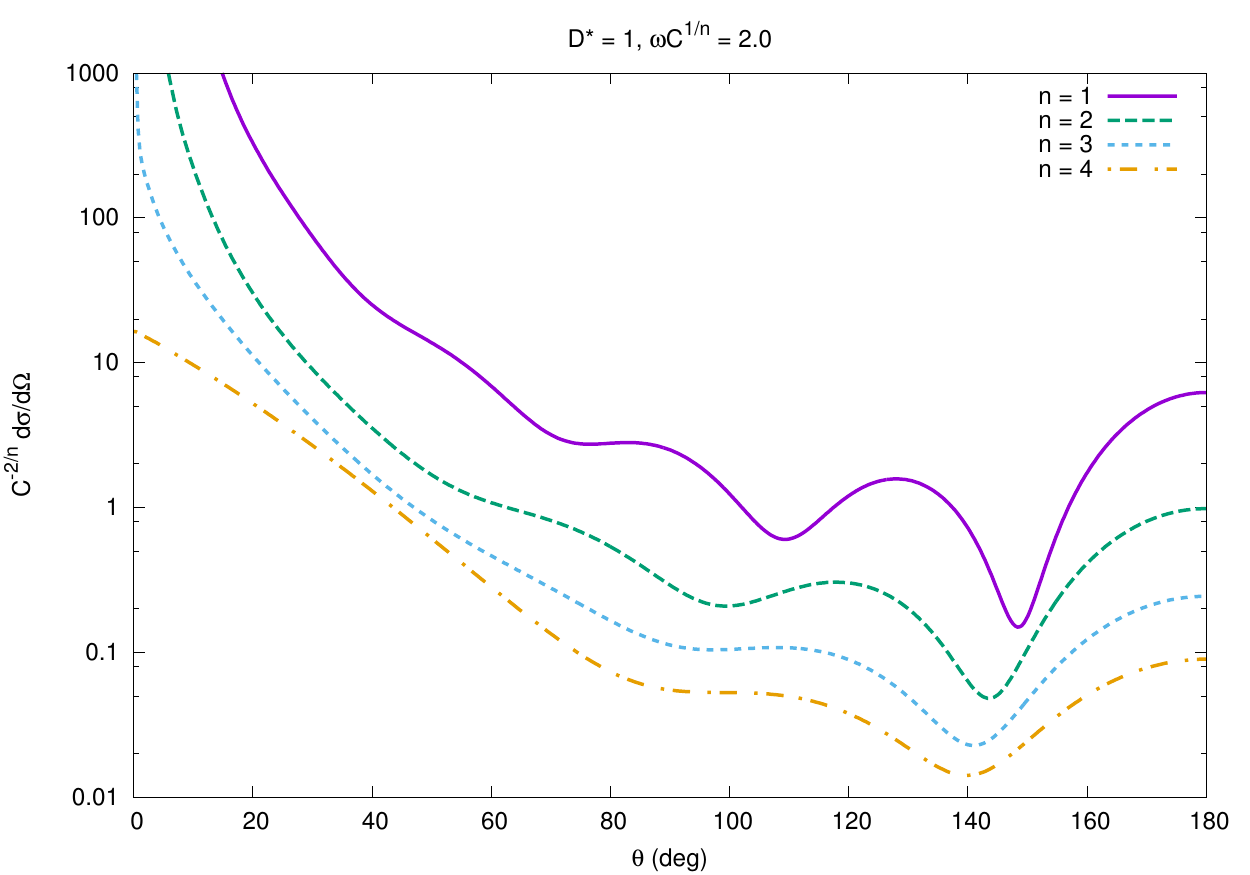}
\caption{Comparison of the scattering cross sections for extremely charged black holes in spacetimes with different number of dimensions. Here we consider $\omega C^{1/n} = 2.0$.}
\label{fig:comp_scs_n}
\end{figure}

Figure~\ref{fig:comp_scs_n} shows a comparison of the scattering cross sections for extremely charged black holes in spacetimes with different dimensions, namely $n = 1, 2, 3, 4$ (case $n = 1$ has been extensively studied in Ref.~\citep{crispino2009prd79_064022}). Similarly to what happens with uncharged black holes~\citep{marinho2016arxiv1612_05604}, the increase in the number of dimensions implies in weaker interactions between the black hole and the test field. As a consequence, the intensity of scattered flux by the black hole decreases as $n$ increases. The fringes of interference get wider with the increase of $n$. This is in agreement with the results presented in Tab.~\ref{tab:gl_params}, where we observed a decrease of $b_g$ with the increase of $n$.

\section{Final remarks}\label{sec:conc}

We have computed the differential scattering cross section for the massless scalar field localized on the 3-brane of higher-dimensional charged black holes in the ADD model. We have focused mainly in the case of extreme black holes, but also presented results for non-extreme and uncharged black holes. The latter case has been extensively studied in Ref.~\citep{marinho2016arxiv1612_05604}. Although we showed results only for the cases $n = 1,2,3,4$ our analysis could be straightforwardly generalized for higher values of $n$.

We would like to remark the fact that there is an important transition in the behavior of the cross section when changing the spacetime dimensionality from $n < 4$ to $n \ge 4$. For the cases $n \ge 4$, the differential scattering cross section is finite in all directions, while it diverges in the forward direction for the cases $n < 4$. This has been foreseen based on the Born approximation in Ref.~\citep{marinho2016arxiv1612_05604} for uncharged black holes, but the conclusions there can be equally applied to the cases studied here since the effects of the black hole charge can be neglected in the weak-field limit. This has been evidenced in the results obtained via geodesic (Fig.~\ref{fig:geo_sca}) and partial-wave (Fig.~\ref{fig:charge_comp}) methods where we have observed that the curves which describe the differential scattering cross sections for black holes with different charges tend to approach each other with the decrease of $\theta$, being almost indistinguishable at $\theta \lesssim 40^\circ$.

The results obtained via partial-wave method where compared with two approximations: (i) the geodesic approximation and (ii) the glory approximation. Both comparisons resulted in excellent agreements in their regime of validity for the cases $n \le 3$; small angles in case (i) and large angles in case (ii). For $n = 4$, the agreement is excellent in case (ii) but not in case (i) since the classical approximation predicts a divergence in the differential scattering cross section in the limit $\theta \to 0$, while the wave analysis reveals that the scattered flux is actually finite in such limit. The same disagreement between the partial-wave method and the geodesic approximation is expected to happen in the cases $n > 4$, unless one considers $\omega C^{1/n} \to \infty$.

\section*{Acknowledgments}

The author would like to thank to Conselho Nacional de Desenvolvimento Cient\'ifico e Tecnol\'ogico (CNPq) and Coordena\c{c}\~ao de Aperfei\c{c}oamento de Pessoal de N\'ivel Superior (CAPES) for partial financial support.
%

\end{document}